\documentclass[12pt]{article}

\usepackage{cite}
\usepackage{subfigure}
\usepackage{multirow}
\usepackage{helvet}
\usepackage{amsmath}
\usepackage{amssymb}
\usepackage{setspace}
\usepackage{setspace}
\usepackage[dvips]{graphicx}
\usepackage{epsfig}

\begin{document}

\begin{titlepage}
\begin{flushright}
IPPP/12/99, DCTP/12/198, Cavendish-HEP-12/20
\end{flushright}
\vspace{0.1cm}

\begin{center}
{\Large\bf Latest Results in Central Exclusive\\[.2cm] Production: A Summary}\\[1.3cm]
{\large {}\footnote{KRYSTHAL collaboration}L.A. Harland--Lang$^a$, V.A. Khoze$^a$, M.G. Ryskin$^b$\\[.2cm] and W.J. Stirling$^c$}\\[1cm]
$^a$ Department of Physics and Institute for Particle Physics Phenomenology, University of Durham, DH1 3LE, UK\\[7pt]
$^b$ Petersburg Nuclear Physics Institute, NRC Kurchatov Institute, Gatchina, St. Petersburg, 188300, Russia\\[7pt]
$^c$ Cavendish Laboratory, University of Cambridge, J.J.\ Thomson Avenue, Cambridge, CB3 0HE, UK
\end{center}

\begin{center}
\begin{abstract}
\noindent Selected new results in central exclusive production (CEP) 
processes within the pQCD--based Durham model are discussed\footnote{Based on talk given by V.A. Khoze at `Diffraction 2012' Workshop,
Puerto del Carmen, Lanzarote, Spain, Sept. 10--15th, 2012.}. Topics covered include the CEP of SM and BSM Higgs--like particles, meson pair CEP and the gap survival probability.
\end{abstract}
\end{center}


\end{titlepage}

\section{Introduction}

\noindent There has recently been a rise in interest in studies of CEP
processes in high--energy proton--(anti)proton collisions, both theoretically and experimentally, 
see e.g.~\cite{Martin:2009ku,Albrow:2010yb,Ryutin:2012np,albrow,royon,Lebiedowicz,chytka,Li}. The CEP of an object $X$ may be written in the form
\begin{equation}\label{cep}
pp(\overline{p}) \to p+X+p(\overline{p})\;,
\end{equation}
where $+$ signs are used to denote the presence of large rapidity gaps. An important advantage
 of these reactions is that they provide an especially clean environment in which to probe
  the nature and quantum numbers of new resonance states,
   from `old' SM mesons to BSM Higgs--like particles (see, for instance ~\cite{Khoze:2010ba}, \cite{HarlandLang:2010ep,Heinemeyer:2007tu,Kaidalov:2003fw,Pasechnik:2007hm}).
One important example is the CEP of the Higgs boson, 
which is at the heart of the FP420 LHC  project~\cite{Albrow:2008pn},
proposing to complement the ATLAS and CMS experiments
by  additional near--beam proton detectors  420m away from the interaction region.
This subject has become
especially topical nowadays, after the LHC discovery 
of a new $\sim$126 GeV Higgs--like boson~\cite{DeRoeck}.
The forward proton technique is
 exceptionally well suited for the investigation
of crucial identification issues such as the spin and $CP$ parity
and the $b\overline{b}$ coupling of the newly discovered object.
This approach is 
complementary to the mainstream strategies at the LHC, and could be
useful for the studies of heavier Higgs--like particles
expected in BSM theories~\cite{Heinemeyer:2007tu}.
It is worth recalling that the observation of even a few events
corresponding to the CEP of a Higgs--like particle would
confirm its $0^{++}$ nature, with the $0^{-+}$, $2^{-+}$
and $2^{++}$ (in the case of minimal coupling to gluons) assignments being strongly disfavoured~\cite{Kaidalov:2003fw,Khoze:2001xm,HarlandLang:2009qe,Khoze:2002nf}.

Note also that the correlations
between the outgoing proton momenta in the CEP mode
would provide a unique possibility to hunt for $CP$--violation
effects in the Higgs sector~\cite{Khoze:2004rc}. 
A  promising program of QCD and new physics studies is under discussion in the
framework of the AFP~\cite{royon,chytka} and HPS, Stage 1~\cite{albrow}
upgrade projects, which would allow an investigation of the region of centrally produced masses
around 200--800 GeV, using proton detectors stationed at $\sim$220m  and $\sim$240m
from the interaction points of ATLAS and CMS, respectively.

As discussed in~\cite{HarlandLang:2010ep,HarlandLang:2011qd},
 the CEP of, for instance, $\gamma\gamma$, dijets, heavy $(c,b)$ quarkonia,
  new charmonium--like states or meson pairs with sufficiently large $p_\perp$ can serve
   as `standard candle' processes to benchmark predictions for new CEP physics,
  as well as offering a promising way to study various aspects of QCD. 
 The expected cross sections and final--state particle distributions 
 are  determined by a non--trivial convolution of the hard amplitude $T$ and the so--called soft survival factors $S^2$, 
defining the probability that the rapidity gaps survive soft and semi--hard rescattering 
effects~\cite{Martin:2009ku,martin}.
This is modeled in the SuperCHIC Monte Carlo~\cite{SuperCHIC}, which allows for an exact generation on an event--by--event basis of the  distributions of the final--state  particles.
\begin{figure}
\begin{center}
\includegraphics[scale=1.0]{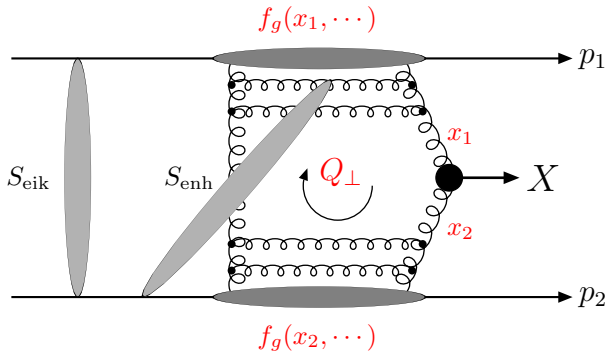}
\caption{The perturbative mechanism for the exclusive process $pp \to p\,+\, X \, +\, p$, with the eikonal and enhanced survival factors 
shown symbolically.}
\label{fig:pCp}
\end{center}
\end{figure} 

In 2007 CDF published a search for $\gamma\gamma$ CEP~\cite{Aaltonen:2007am} at the Tevatron, with $E_\perp(\gamma) >$ 5~GeV. Three candidate events were observed, in agreement with the expectation of~\cite{Khoze:2004ak}. Subsequently, to increase statistics the $E_\perp(\gamma)$ threshold has been decreased to 2.5 GeV, and in~\cite{Aaltonen:2011hi} the  observation of 43  $\gamma\gamma$ events in $|\eta(\gamma)|<1.0$ with no other particles detected in $-7.4<\eta<7.4$ was reported, which corresponds to a cross section of $\sigma_{\gamma\gamma}= 2.48^{+0.40}_{-0.35} $ $({\rm stat})^{+0.40}_{-0.51}$ $ ({\rm syst})$ pb. The theoretical cross section, calculated using the formalism described in~\cite{HarlandLang:2010ep,Khoze:2004ak} and implemented in the SuperCHIC MC generator~\cite{SuperCHIC}, is 1.42 pb using MSTW08LO PDFs~\cite{Martin:2009iq} and 0.35 pb using MRST99 (NLO) PDFs~\cite{Martin:1999ww} (a comparison with a wider range of PDF sets is made in~\cite{HarlandLang:2012qz}), while the $p_\perp$, $\Delta \phi$ and invariant mass distributions of the $\gamma\gamma$ pair are well described by the MC.

The LHCb Collaboration has reported preliminary
 results~\cite{LHCb} on the CEP of $\chi_{c}$ mesons in the $\chi_c\to J/\psi\,+\,\gamma$ channel
 with vetoing on additional activity in the rapidity region $1.9<\eta<4.9$, 
and some sensitivity to charged particles in the backwards region $-4<\eta<-1.5$~\cite{LHCb}.
 While the $\chi_{c(0,1)}$  data are in good agreement with the CEP predictions, 
 the observed $\chi_{c2}$ rate is somewhat higher. However, as discussed in~\cite{HarlandLang:2011qd},
the  LHCb data include a contribution of events with proton dissociation, which
 favour the production of higher spin $\chi_{c(1,2)}$ states, with the $\chi_{c2}$ yield being particularly enhanced.
However a more accurate account of the effects caused by the un--instrumented
regions in the LHCb experiment~\cite{LHCb} requires more detailed
quantitative studies. In~\cite{LHCb} a cut of $p_\perp<0.9$ GeV on the $\mu^+\mu^-$ system is placed, which will reduce the contribution from in particular higher mass proton dissociation (i.e. with larger $k_\perp $ transferred through the $t$--channel). Such contamination would be expected to particularly enhance the $\chi_{c2}$ cross section, which we recall increases as $\sim \langle p_\perp^2 \rangle^2$. As discussed in~\cite{HarlandLang:2012qz}, with the higher statistics that will come with future data, a detailed study of the $p_\perp(\chi_c)$ dependence of the cross section ratios $\sigma(\chi_{c(1,2)})/\sigma(\chi_{c0})$ would shed important light on this. As the $p_\perp$ of the central system decreases, any contamination from events with proton dissociation should decrease and we would expect the cross section ratios to become more consistent with the exclusive predictions.

Another way to clarify the situation (see~\cite{Lebiedowicz,HarlandLang:2010ep,HarlandLang:2009qe,Khoze04}) 
is to consider other decay modes, for instance  the observation of $\chi_{c0}$ CEP via two--body 
channels ($\pi^+\pi^-$, $K^+K^-$, $p\overline{p}$,...). Considering the case of $\chi_c \to \pi^+\pi^-$ CEP for example, while the $\chi_{c0}$ cross section
 is of the same size as in the $\chi_{c0} \to J/\psi \gamma \to \mu^+\mu^- \gamma$ channel,
 the fact that the $\chi_{c(1,2)}$ two--body  branching ratios are smaller
 (or even absent for the $\chi_{c1}$) than for the $\chi_{c0}$, ensures that the $J_z=0$ selection rule~\cite{Khoze:2001xm,Khoze:2000jm} is fully active.
 However, here we may expect a sizeable background  from direct QCD $\pi^+\pi^-$ production. This process can be modeled in two different ways: for low invariant mass and/or transverse momentum final states a `non--perturbative' mechanism, calculated using the tools of Regge theory, should dominate, while 
the high $k_\perp$ tail of the $\pi^+\pi^-$ CEP process should be generated by a purely pQCD--mechanism. We shall consider both of these mechanisms in more detail below.

\section{Meson pair production}

\subsection{Non--perturbative CEP mechanism}

\begin{figure}[h]
\begin{center}
\includegraphics[scale=0.9]{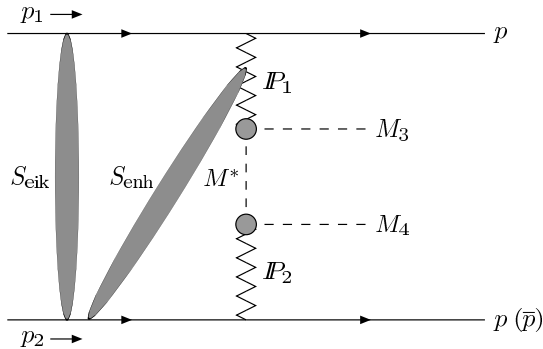}
\caption{Representative diagram for the non--perturbative meson pair ($M_3$, $M_4$) CEP mechanism, where $M^*$ is an intermediate off--shell meson of type $M$. Eikonal and (an example of) enhanced screening effects are indicated by the shaded areas.}\label{npip}
\end{center}
\end{figure}

For the non--perturbative contribution we expect a picture of the type shown in Fig.~\ref{npip} to dominate.
 For such a mechanism, the meson ($\pi^+\pi^-$...) pair is created via double--Pomeron exchange, with an intermediate $t$--channel off--shell meson. The amplitude is calculated in~\cite{HarlandLang:2011qd}. The CEP matrix 
element is given by $\mathcal{M}=\mathcal{M}_{\hat{t}}+\mathcal{M}_{\hat{u}}$, with $\hat{t}=(P_1-k_3)^2$, $\hat{u}=(P_1-k_4)^2$, where $P_i$ is the momentum transfer through Pomeron $i$, and $k_{3,4}$ are the meson momenta. We have
\begin{equation}\label{namp}
\mathcal{M}_{\hat{t}}=\frac 1{M^2-\hat{t}} F_p(t_1)F_p(t_2)F^2_M(\hat{t})\sigma_0^2
\bigg(\frac{s_{13}}{s_0}\bigg)^{\alpha(p_{1\perp}^2)}\bigg(\frac{s_{24}}{s_0}\bigg)^{\alpha(p_{2\perp}^2)}\;,
\end{equation}
where $M$ is the meson mass and $s_{ij}=(p_i'+k_j)^2$ is the c.m.s. energy squared 
of the final state proton--meson system $(ij)$. 
In the $\pi^+\pi^-$ case the normalisation is set by the total pion--proton cross section
 $\sigma(\pi p)=\sigma_0 (s_{ij}/s_0)^{\alpha(0)-1}$. The $F_M(\hat{t})$ in (\ref{namp}) is the 
form factor of the intermediate off--shell meson and, as discussed in~\cite{HarlandLang:2011qd}, 
it is quite poorly known, in particular for larger values of $\hat{t}$. Traditionally, 
 a typical `soft' exponential form is taken $F_M(\hat{t})=\exp{(b_{\rm off}(\hat{t}-M^2)})$,
 and the value of the slope is approximately fitted to reproduce the correct normalisation of CERN--ISR data.
It is worth mentioning that
 the $t$--channel state $M^*$ could correspond not only to pion exchange but also
 to the exchange of heavier states ($a_1,\ a_2,...$), which could modify the CEP cross section
at moderate  pion $k_\perp$. Such effects, in particular, $a_2$ exchange,
may have already revealed themselves in the new CDF
measurement of the dipion mass distribution at 
 $M_{\pi\pi}<5.5$ GeV at 900 GeV and 1.96 TeV~\cite{albrow}.
Currently there is no deep theoretical understanding concerning 
the form of $F_M(\hat{t})$, which appears to be the `Achilles  heel' of such a 
non--perturbative  model. 
In order to have a better sensitivity to this  form factor, the meson
$k_\perp$-- distributions corresponding to the same data would be very useful.
Also 
a comparison between the $k_\perp$ ($M_{\pi\pi}$) distributions
at 1.96 TeV and 900 GeV would probe the size of any possible contamination due to proton dissociation.

Finally, we recall that to calculate the genuine CEP cross section we have to include an additional suppression, accounting for screening corrections, that is the eikonal survival factor, $S_{\rm eik}$, and the {\em enhanced} survival factor, $S_{\rm enh}$, depicted in Fig.~\ref{npip} in terms of Pomeron exchanges. Following~\cite{HarlandLang:2011qd}, we also introduce an extra suppression of the form of $\exp(-n)$, corresponding to the small Poisson probability not to emit other secondaries in the $I\!\!P I\!\!P\to M_3\overline{M}_4$ process, where $n(s_{M\overline{M}})$ is the mean number of secondaries.

\subsection{Perturbative CEP mechanism}

\begin{figure}
\begin{center}
\includegraphics[scale=1.4]{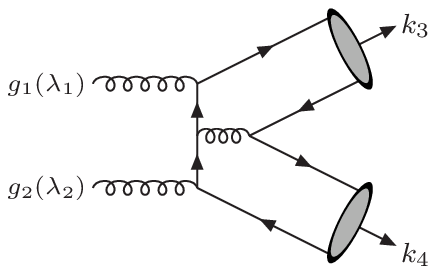}\qquad\qquad
\includegraphics[clip,trim=-20 0 0 0,scale=0.8]{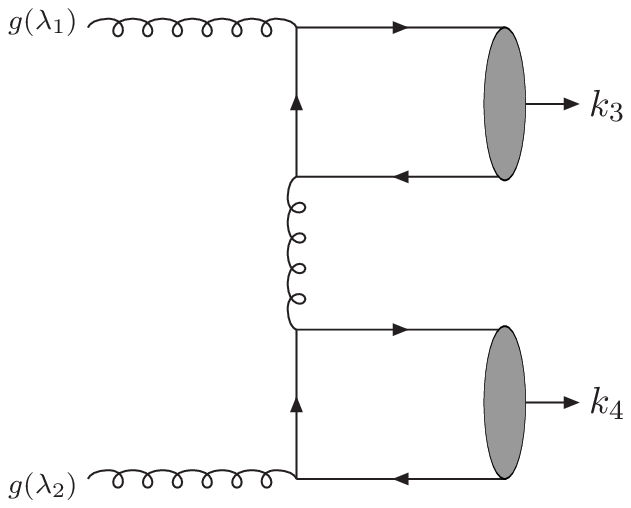}
\caption{(Left) A typical diagram for the $gg \to M\overline{M}$ process. (Right) Representative `ladder' diagram, which contributes to the production of flavour--singlet mesons.}\label{ladder}
\end{center}
\end{figure}

At higher values of the meson $k_\perp$ we  model the meson pair CEP process using the pQCD--based Durham model, 
as in Fig.~\ref{fig:pCp}. To calculate the $gg \to M\overline{M}$ subamplitude we generalise the `hard exclusive' formalism used to calculate the $\gamma\gamma \to M\overline{M}$ cross section~\cite{Brodsky81,Benayoun89}. 
We then calculate the relevant parton--level helicity amplitudes for the $gg \to M\overline{M}$ process, for the production of scalar flavour--nonsinglet meson states ($\pi\pi$, $K^+K^-$, $K^0\overline{K}^0$). There are seven independent Feynman diagrams to compute -- a representative diagram is given in Fig.~\ref{ladder} (left). 
The results of explicit calculations are given in~\cite{HarlandLang:2011qd}.
The $gg\to M\overline{M}$ amplitude has a remarkable property:
 for $J_z=0$ gluons it vanishes at LO for scalar flavour--nonsinglet mesons, which, recalling 
the $J_z=0$ selection rule~\cite{Khoze:2001xm,Khoze:2000jm} that strongly suppresses the CEP of non--$J_z=0$ states, 
will lead to a strong suppression (by $\sim$ two orders of magnitude) in the CEP cross section. 
As a result, we may expect the perturbative contribution to the continuum background to $\chi_c \to \pi^+\pi^-$ to be small.
It is also found~\cite{HarlandLang:2011qd}, that the $|J_z|=2$ amplitude is additionally suppressed by the presence of a `radiation zero' at a particular value of the scattering angle.
An  important consequence of this
is that the $\pi^0\pi^0$ QCD background to the $\gamma\gamma$ CEP process is predicted to be small. 
As discussed above, in~\cite{Aaltonen:2011hi} CDF reported the  observation of 43 $\gamma\gamma$ events with $|\eta(\gamma)|<1.0$ and  $E_T(\gamma)>2.5$ GeV. In this analysis  special attention was paid to the possible background caused by $\pi^0\pi^0$ CEP, since  the photons 
from $\pi^0 \to \gamma\gamma$ decay can mimic the `prompt' photons from $gg \to \gamma\gamma$ CEP. 
Importantly, CDF has found that the contamination caused by $\pi^0\pi^0$   CEP is very small 
 (corresponding to a ratio $N(\pi^0\pi^0)/N(\gamma\gamma)<0.35$, at 95\% CL), supporting this result (which predicts $N(\pi^0\pi^0)/N(\gamma\gamma)\sim 1\%$).
The first CMS results on $\gamma\gamma$ CEP will be available soon~\cite{Li}.

It is also possible for the $q\overline{q}$ forming the mesons to be connected by a quark line, as shown in Fig.~\ref{ladder} (right). These will only give a non--zero contribution for the production of ${\rm SU}(3)_F$ flavour--singlet states, 
i.e. $\eta'\eta'$ and, through $\eta$--$\eta'$ mixing, $\eta\eta$ and $\eta\eta'$ production.
 The explicit amplitudes are given in~\cite{HarlandLang:2011qd}, but the crucial 
result is that the $J_z=0$ amplitudes do not vanish as in the case 
of flavour non--singlet mesons, and so we will expect $\eta'\eta'$ CEP
 to be strongly enhanced relative to, for example, $\pi\pi$ production, due to the $J_z=0$ selection 
rule which operates for CEP.
 In the case of $\eta\eta$ production, the flavour singlet contribution 
will be suppressed by a factor $\sin^4 \theta_P$, where $\theta_P$ 
is the octet--singlet mixing angle, and this may therefore
 be comparable to the $|J_z|=2$ flavour--octet contribution. In fact, after an explicit
 calculation we find that the $\eta\eta$ CEP cross section is expected, in the regions where the 
perturbative formalism is applicable,  to be dominant over $\pi\pi$ CEP. 

It is also worth mentioning that studies of $\eta'\eta'$, $\eta'\eta$ and $\eta\eta$ CEP could 
provide unique information about the gluonic component of the $\eta'$ (and $\eta$) mesons. In particular, any $gg$ component of these mesons will contribute to the CEP process via diagrams as in for example Fig.~\ref{ladder}, but with one or both quark lines replaced by gluon pairs with the correct quantum numbers to form the $\eta(\eta')$ state. Indeed, we find by explicitly calculating the corresponding $gg \to gggg$ and $gg\to ggq\overline{q}$ amplitudes~\cite{fut} that the $\eta'\eta'$, $\eta'\eta$ and $\eta\eta$ CEP cross sections display significant sensitivity to the size of this $gg$ component. As discussed in~\cite{HarlandLang:2012qz}, there is uncertainty regarding the size of such a component, and so the observation of these CEP processes could shed some important light on the issue. We may expect new results on this to come from further analysis of the existing CDF data as well as from the CMS/Totem (ATLAS) special low--luminosity runs~\cite{albrow}. 

\section{Exotic charmonium--like states: a comment on the CEP of the $X(3872)$}

The CEP mechanism could also provide a complementary way to shed light on the nature of the large number of `exotic' XYZ charmonium--like states which have been discovered over the past 10 years, see for example~\cite{Brambilla:2010cs} for a review. In some cases the $J^{PC}$ quantum numbers of these states has not been determined experimentally, and often a range of interpretations are available for these states: a $D^0\overline{D}^{*0}$ molecule, tetraquarks, $c\overline{c}g$ hybrids, the conventional $c\overline{c}$ charmonium assignment, and more generally a mixture of these different possibilities. Considering in general the CEP of such objects, then the $J_z^{P}=0^{+}$ selection rule which~\cite{Khoze:2001xm,Khoze:2000jm} strongly suppresses the CEP of non--$J_z^{P}=0^{+}$ states, as well as a measurment of the distribution  in the azimuthal angle $\phi$ between the transverse momenta of the outgoing protons (as in e.g.~\cite{Kaidalov:2003fw,HarlandLang:2010ep}), may help to fix the quantum numbers of the centrally produced system. Moreover, since the original $c\overline{c}$ pair is in this case produced at rather short distances, the CEP process can probe the wavefunction of the corresponding charmonium at the origin.

The well--known $X(3872)$ is a particularly interesting example of this, as it was the first such state to be discovered (by BELLE in 2003~\cite{Choi:2003ue}), with a concrete interpretation  for it still remaining elusive. It has become even more topical with the recent establishment of its quantum numbers to be $J^{PC}=1^{++}$ by LHCb~\cite{Aaij:2013zoa}, an assignment which leaves both the more exotic and the conventional $\chi_{c1}(2{}^3 P_1)$ interpretations in principle available, as well a combination of, for example, the $c\overline{c}$ and molecular $D$ meson states, see for example~\cite{Voloshin:2007dx} for a review.

The $X(3872)$ has been seen in prompt inclusive production at both the Tevatron and LHC, and this raises the interesting possibility of observing its production in the exclusive channel. Such an observation would first of all probe the direct (i.e. not due to feed--down from the decay of higher mass states) production channel $gg \to X$ of this state. If the $X(3872)$ is a $D^0\overline{D}^{*0}$ molecule, then the binding energy of this would have to be very small, and so such a loosely bound system would have to be produced with a very small relative $k_\perp$ in the $D^0\overline{D}^{*0}$ rest frame, corresponding to a large separation between the mesons. As discussed in~\cite{Artoisenet:2009wk,Bignamini:2009sk}, the hadroproduction of such a state with the size of cross section observed in the $X(3872)$ case (see e.g.~\cite{Chatrchyan:2013cld}), if possible at all, should in general take place in an environment where additional particles are emitted, so that the initially produced short--distance $c\overline{c}$ pair can form a loosely--bound, $D^0\overline{D}^{*0}$ state, at long distances. We would expect such a transition to be quite rare in the exclusive case, where no additional particles can be present, and so the observation of $X(3872)$ CEP would on general grounds disfavor such a purely molecular interpretation.

For a conventional $\chi_{c1}(2{}^3 P_1)$ state, the ratio of the CEP cross sections $\sigma(\chi_{c1}(2P))/\sigma(\chi_{c1}(1P))$ is predicted to first approximation (ignoring reasonably small corrections due to the different masses, relativistic effects etc) to be simply given by the ratio of the respective squared wavefunctions at the origin $|\phi_P'(0)|^2$. That is, we will expect them to be of comparable sizes. Moreover, we should recall that the CEP of the ground--state $\chi_{c1}(1P)$ has already been observed by LHCb~\cite{LHCb}, thus raising the possibility of such a measurement in the same experimental conditions. This result of course depends on the conventional charmonium interpretation for the $X(3872)$ being valid. If, as may be more realistic, it is a mixture of a $\chi_{c1}(2P)$ and a molecular $D^0\overline{D}^{*0}$ state, then the size of this ratio will also  be driven by the probability weight of the purely $c\overline{c}$ component; if this is small, that is the molecular component is dominant, then the $X(3872)$ cross section will be suppressed relative the to the $\chi_{c1}(1P)$: as discussed above we would not expect the molecular component to be accessed in an exclusive environment. In this way, the CEP mechanism could shed light on the nature of this puzzling state.

\section{Gap survival probability}\label{gapsec}

The soft survival factor $S^2$ plays a crucial role in the evaluation of the rate of CEP processes, but to compute this suppression factor we have to apply a given model of soft hadron scattering, see for example~\cite{Martin:2009ku,martin}. Before the Totem data on elastic proton--proton scattering at 7 TeV~\cite{mario} became available a value of $S^2\sim 0.02$ at 14 TeV was preferred by the Durham and other theory groups, see e.g.~\cite{Martin:2009ku,Ryskin:2009tk} and~\cite{Gotsman:2012rm} for discussion and further references. However, these model expectations underestimated the total proton--proton cross section, $\sigma_t$, at 7 TeV, as measured by Totem~\cite{mario}. Since the probability of additional interactions is proportional to $\sigma_t$, a larger value for this will lead to a smaller gap survival probability $S^2$.  

There are two types of additional inelastic interactions: the rescattering of the incoming soft parton spectators, and the interaction of the intermediate partons, created in the evolution of one proton with another proton. The corresponding probabilities of gap survival are denoted by $S^2_{\rm eik}$ and $S^2_{\rm enh}$, respectively. In the latter case the partons in the evolution have a relatively large transverse momenta, $k_t$, and correspondingly a relatively small ($\propto 1/k^2_t$) absorptive cross section. The main suppression is therefore provided by inelastic interactions of the parton spectators, and so here we will only consider the case of `eikonal' screening $S^2_{eik}$ (for more details of the calculation of the `enhanced' survival factor $S_{\rm enh}$, see e.g.~\cite{Ryskin:2009tk,Ryskin:2011qe} and references therein).

In the one--channel eikonal model the probability of no additional interactions is directly related to the elastic scattering amplitude. In impact parameter, $b$, space 
\begin{equation}\label{s1}
S_{\rm eik}=|1+iA(b)|=\exp(-\Omega(b)/2), 
\end{equation}
where the `partial wave' elastic amplitude is given by $A(b)=i\left(1-\exp(2i\delta(b))\right)$ and $\Omega=4{\rm Im}\delta(b)$. Neglecting the small ($\rho={\rm Re}A/{\rm Im}A\ll 1$) real part of the elastic amplitude, the function $A(b)$ can be obtained from experimental data
\begin{equation}\label{a}
A(b)=i\int\sqrt{\frac{{\rm d}\sigma}{{\rm d}t}\frac{16\pi}{1+\rho^2}}
e^{i\vec q_t\cdot\vec b}\frac{{\rm d}^2q_t}{8\pi^2}\simeq i\int\sqrt{\frac{{\rm d}\sigma}{{\rm d}t}}
J_0(q_t b)\frac{q_tdq_t}{\sqrt\pi}\; ,
\end{equation}
where $t=-q_t^2$, $J_0$ is a Bessel function and we normalized so that $A=i$ corresponds to the black disk limit of the elastic amplitude.

At the first sight it appears that (\ref{s1},\ref{a}) define the value of $S_{\rm eik}$ unambiguously. However, the problem is that such a model only accounts for the proton in the intermediate state, and neglects the possibility of $p\to N^*$ excitations. Moreover, data from old fixed target experiments indicate that the probability of `quasi--elastic' $pp\to pN^*$ processes is as large as 30\%~\cite{Kaid}. To allow for these $p\to N^*$ transitions we have to use the Good--Walker formalism~\cite{GW}, decomposing the incoming proton wave function into a number of eigenstates $|p\rangle=\sum a_i|\phi_i\rangle$ which diagonalize the amplitude $A_{ik}=\beta_i\delta_{ik}$. Each eigenstate $|\phi_i\rangle$ has its own interaction cross section, $\sigma_i\propto \beta_i$, and its own radius $R_i$. As a rule, the existing models include two or three such diffractive eigenstates.
  
An open question is how the partons in the incoming proton wave function are distributed between these states $|\phi_i\rangle$.
Global parton analyses only probe the distribution summed over all the states, $f(x)=\sum f_i(x)|a_i|^2$, but not the distributions, $f_i(x)$, in each individual state. The situation was not so crucial at lower energies, when the transparency of the disk, that is $S_{\rm eik}=1+iA(b)$, was relatively large. But now the Totem data indicate that at small impact parameters, $b\to 0$, the elastic amplitude has already reached the black disk limit ($A=i$), and in this case the resulting gap survival factor, $S^2_{\rm eik}$ becomes very sensitive to the distribution of the partons between the different Good--Walker eigenstates.

It appears to be most natural to assume that the parton density in a state $|\phi_i\rangle$ is proportional to its cross section, that is to $\beta_i$. However we can not exclude the possibility that, for example, gluons are mainly concentrated in the state with the largest cross section (largest $\beta_i$) while the quarks are concentrated in the state with the smaller $\beta_i$. In the interaction of two eigenstates $|\phi_1\rangle$ with the largest cross section, saturation may be reached up to a rather large $b\sim 0.7$ fm, that is almost the whole disk becomes `black' with $S^2_{11}=0$, while for the two smallest eigenstates, say $|\phi_3\rangle$, saturation is only reached in the centre, $b\lesssim 0.1$ fm, with $S^2_{33}\simeq 1$ over most of the impact parameter range ($b\gtrsim 0.2$ fm).

Assuming that the parton distributions are proportional to the optical density, $f_i\cdot f_k\propto \Omega_{ik}(b)$ and using the model of~\cite{Opacity} it is found that at $\sqrt s=14$ TeV the survival factor\footnote{Throughout this note, when quoting such values for the soft suppression factors we are more precisely referring to the value of the survival factor {\em averaged} over the outgoing proton $p_\perp$, i.e. $\langle S^2_{\rm eik}\rangle$, see e.g.~\cite{HarlandLang:2010ep}. Similarly in the case of the $S^2_{\rm enh}$ it is the survival factor averaged over the parton $k_\perp$ that is relevant.} $S^2_{\rm eik}=0.005$. However this value could be 2--3 times larger if the gluons are distributed between the $|\phi_i\rangle$ states more homogeneously. It should furthermore be emphasized that currently there is {\em no} model which can describe not only the total and elastic $pp$--cross sections but also the cross sections of low mass proton ($p\to N^*$) dissociation in the LHC energy region 
sufficiently well. Thus at the moment we can only estimate the expected gap survival probability for exclusive Higgs production at $\sqrt{s}=14$ TeV, with $S^2\sim 0.01 - 0.004$.  Clearly we need both a better understanding of the experimental data\footnote{As discussed in~\cite{early}, one of the best ways to monitor the
soft survival factor experimentally is to measure dijet CEP over a wide interval of the jet transverse momentum $p_\perp$. The AFP~\cite{royon,chytka}  and HPS~\cite{albrow} facilities, in combination with central detectors, could provide important experimental information about the behaviour of the soft survival factor.} for proton--proton scattering (including low mass $p\to N^*$ dissociation), and a better theoretical model to describe {\em all} of these data.

\section{Higgs boson CEP}

\begin{figure}
\begin{center}
\includegraphics[scale=0.7]{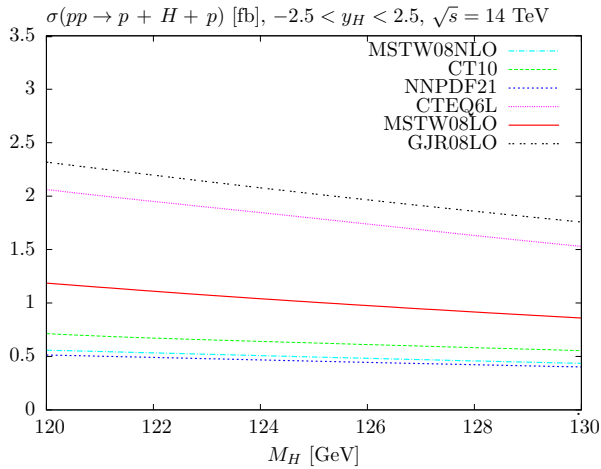}
\caption{Cross section for SM Higgs CEP as a function of the Higgs mass, $M_H$, integrated over the rapidity interval $-2.5<y_H<2.5$. NLO K--factor included.}\label{fig:Hl}
\end{center}
\end{figure}

\begin{figure}
\begin{center}
\includegraphics[scale=0.7]{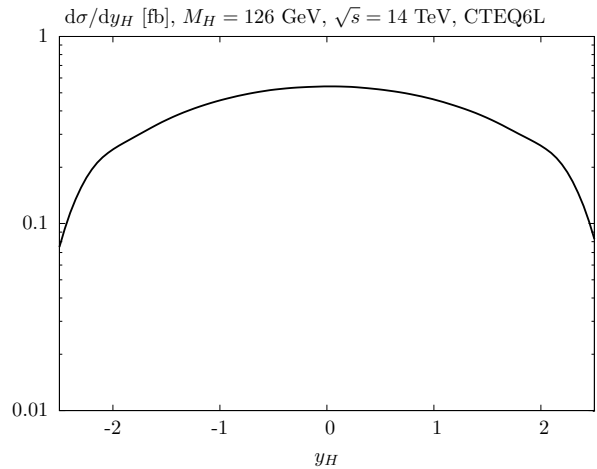}
\caption{Rapidity distribution ${\rm d}\sigma/{\rm d}y_H$ for a $M_H=126$ GeV SM Higgs boson, using CTEQ6L PDFs.}\label{fig:Hr}
\end{center}
\end{figure}

\begin{figure}
\begin{center}
\includegraphics[scale=0.7]{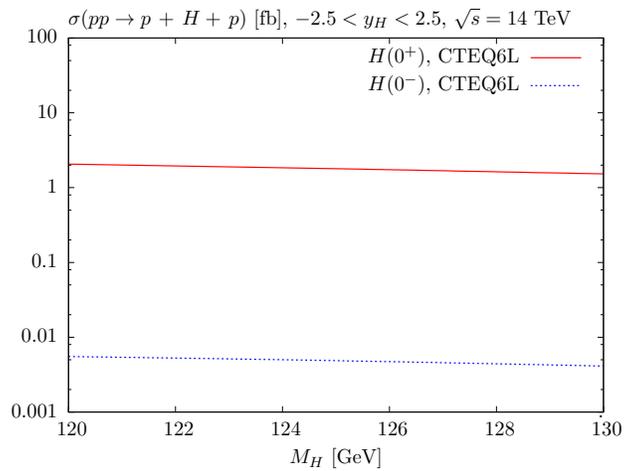}
\caption{Cross sections for the CEP of scalar $J^P=0^+$ and pseudoscalar $J^P=0^-$ particles of the Higgs sector as a function of the Higgs mass, $M_H$, integrated over the rapidity interval $-2.5<y_H<2.5$.}\label{fig:H1}
\end{center}
\end{figure}

The expectations for the CEP of the SM Higgs boson at 14 TeV are illustrated in Figs.~\ref{fig:Hl},~\ref{fig:Hr} and~\ref{fig:H1}. For the combined enhanced\footnote{In this mass and $\sqrt{s}$ region, the suppression due to $S^2_{\rm enh}$ is only expected to be weak~\cite{rmk,Ryskin:2011qe}.} and eikonal soft survival factor we have $S^2=0.01$, although, as discussed above, at $\sqrt{s}=14$ TeV there is an important uncertainty in this value, and it may in particular be somewhat smaller. On the other hand, as discussed in~\cite{HarlandLang:2012qz} we may also expect higher order corrections to increase the cross section by a factor of $\sim 2$ or so. Although there is therefore some important uncertainty in the corresponding estimates for Higgs boson CEP at $\sqrt{s}=14$ TeV, we note that applying the same model with the LO PDFs, which give the larger cross sections in Fig.~\ref{fig:Hl}, there is good agreement with the CDF $\gamma\gamma$ data~\cite{Aaltonen:2011hi}, with the CTEQ6L~\cite{Pumplin:2002vw} 
set giving the closest value. In Fig.~\ref{fig:Hr} we show the corresponding Higgs rapidity distribution for the CTEQ6L PDF set, for $M_H=126$ GeV. In Fig.~\ref{fig:H1} we show the cross section for the case of a scalar $J^P=0^+$ and pseudoscalar $J^P=0^-$ particle of the Higgs sector, using CTEQ6L PDFs. As expected from the $J^P_z=0^+$ selection rule~\cite{Khoze:2001xm,Khoze:2000jm} which operates for CEP, the cross section in the case of the scalar state is much ($\sim 2$ orders of magnitude) larger.

While the predicted scalar Higgs cross sections are quite small ($\sim$ fb), we recall that the CEP process provides an exceptionally clean and complementary handle on the properties of a Higgs or Higgs--like particle. This provides a motivation for the addition of forward proton detectors at 420m, as proposed in the FP420 LHC project~\cite{Albrow:2008pn}, which are essential if such a measurement is to be performed.

\section{A fresh look at the MSSM Higgs CEP}

Though the observed properties of the newly discovered Higgs--like state are in agreement with those of the SM Higgs boson~\cite{DeRoeck}, this spectacular LHC discovery is also compatible with the expectations of the Higgs sector of the MSSM, where the new state could be interpreted as either the light, $h$, or the heavy, $H$, $CP$--even MSSM Higgs boson, while maintaining a SM--like behaviour, see~\cite{Bechtle:2012jw} for details\footnote{VAK thanks Georg Weiglein for a discussion on the interpretation of the  discovered Higgs--like particle  within the MSSM framework.}. Since the MSSM is currently one of the most widely used and well studied BSM scenarios, here we briefly discuss the present expectations for the CEP of the MSSM Higgs bosons, following the approach of~\cite{Heinemeyer:2007tu,Kaidalov:2003ys}.

\begin{figure}
\begin{center}
\includegraphics[scale=0.5]{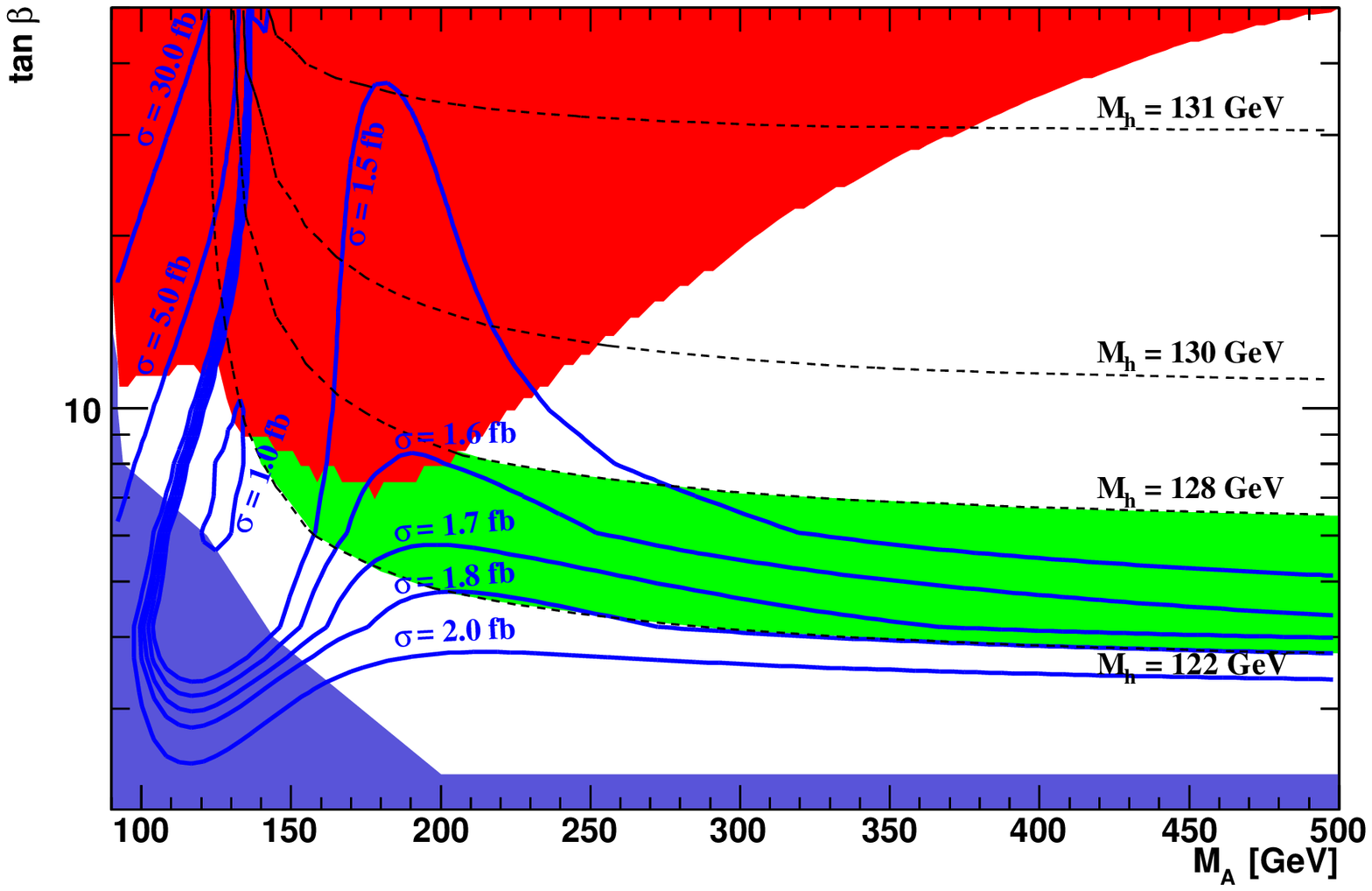}
\includegraphics[scale=0.5]{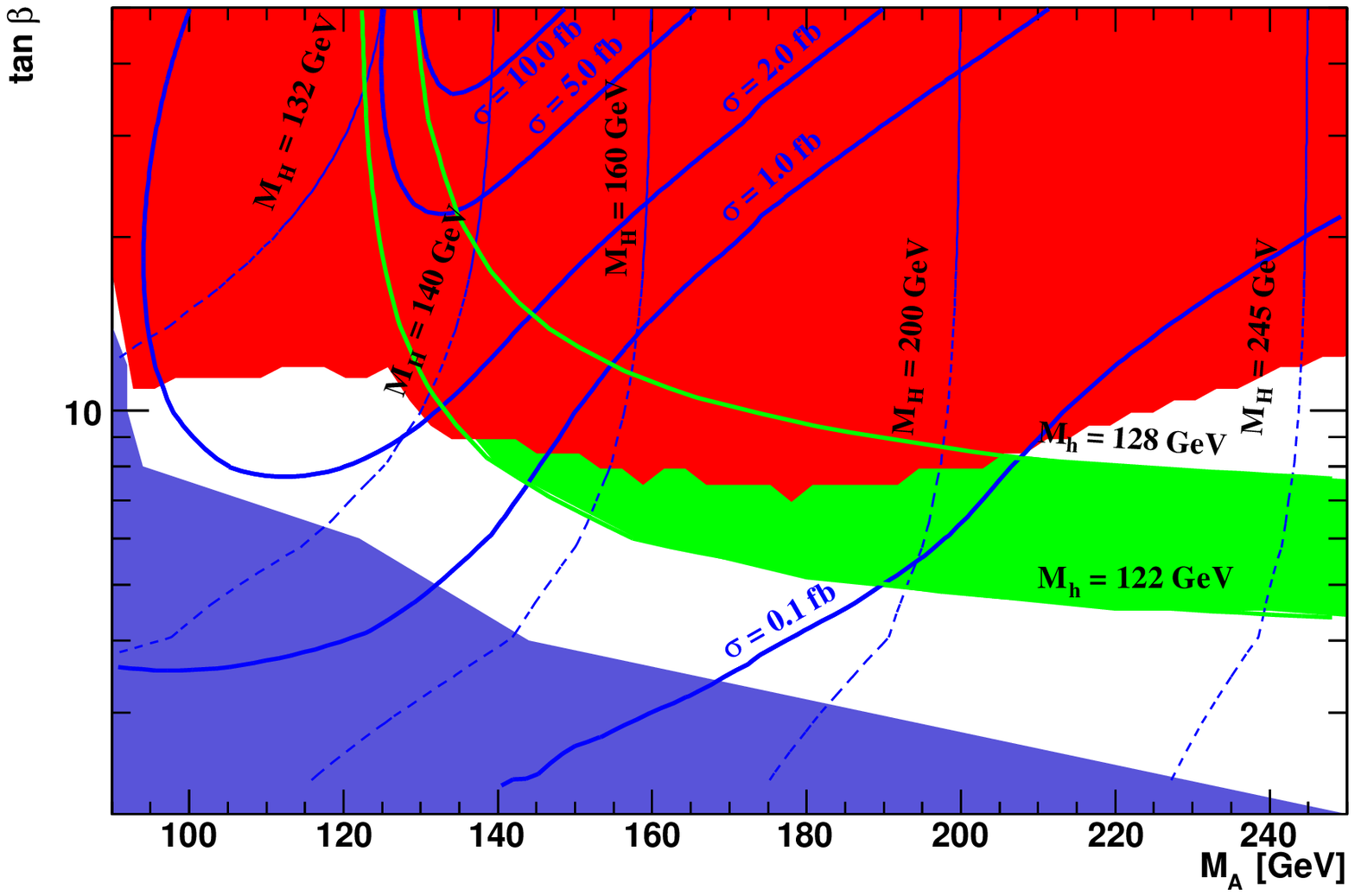}
\caption{Cross sections at 14 TeV for $h(H)\rightarrow b\bar{b}$ CEP in the $M_A-\tan\beta$ plane of the MSSM within the $M_h^{\rm max}$ benchmark scenario~\cite{marek-talks,Heinemeyer:2012hr}. The $h$--boson cross section is shown on the top and the $H$--boson case on the bottom. The blue (red) shaded region corresponds to the parameter region excluded by the LEP (LHC--2011) Higgs boson searches. The 122 $< M_h <$ 128~GeV region corresponds to the identification of the observed 126~GeV mass state as a light $h$ boson, with approximate (and slightly inflated) theory and experimental uncertainties, corresponding to the 2011 LHC data.}

\label{fig:mssm}
\end{center}
\end{figure}

Assuming that the new state is a light MSSM $h$--boson, Fig.~\ref{fig:mssm} shows the results for the CEP of the MSSM Higgs bosons\footnote{Courtesy of Marek Tasevsky, see also~\cite{marek-talks,Heinemeyer:2012hr}. Note that after taking into account the LHC--2012 data the allowed MSSM parameter region will shrink further~\cite{progress}.}. One immediate observation is that the $h$ CEP yield is only weakly dependent on $M_A$, reaching a cross section level of around 1.5 fb, up to a factor of $\sim$ 2 theoretical uncertainty. Thus, after accounting for the new LHC results (contrary to the earlier expectations of~\cite{Kaidalov:2003ys}) the event rate for the MSSM $h$--boson cannot be much higher than that for the SM Higgs. 

The situation for the CEP of the heavy MSSM $H$--boson does not appear to be very promising. Accounting for the recent LHC data and low--energy observables, and again under the assumption that the newly observed state is a light MSSM $h$--boson, the preferred values~\cite{Bechtle:2012jw} of the heavy neutral Higgs masses should be comparatively large (exceeding 250 GeV or so), which is within the acceptance of the 220--240m forward proton detectors~\cite{albrow,chytka}. However, the effective Pomeron--Pomeron luminosity $L^{\rm eff}$ for Higgs boson CEP decreases rapidly with the Higgs mass $M$, being given approximately by~\cite{Khoze:2001xm,Kaidalov:2003ys}
\begin{equation}
L^{\rm eff} \quad \propto \quad 1{\Big /}(M + 16~{\rm GeV})^{3.3}\;.
\label{eqH}
\end{equation}
Including the other mass dependent factors as in~\cite{Heinemeyer:2007tu}, we find that for a $H$--boson mass of 300 (400) GeV the expected cross section is about 0.01 fb (0.001) fb. After accounting for the experimental acceptances and efficiencies in the spirit of~\cite{Heinemeyer:2007tu} (which will reduce the rate by an additional factor of $\sim 0.1$ or so), even with an integrated luminosity of 500--1000 $\mbox{fb}^{-1}$, the event rate will be too low to provide a reasonable significance\footnote{In addition, we have to keep in mind that at higher LHC luminosities the pile--up background could cause a severe problem for CEP measurements, even if/when the fast timing detectors with precision vertex resolution~\cite{albrow,chytka} are installed.}. This conclusion should of course not be generalized to all other BSM Higgs scenarios, some of which might be more favourable.

If we take a more exotic interpretation, where the newly discovered Higgs--like particle is a heavy $CP$--even  MSSM Higgs boson, $H$, as considered, for example in~\cite{Bechtle:2012jw,Drees:2012fb}, then the discovery of the lighter $CP$--even state $h$ could be very challenging at the LHC due to its expected low mass (60--90 GeV) and strongly suppressed coupling to vector bosons. Since the acceptances of the proton detectors of the FP420 project~\cite{Albrow:2008pn} should be well within the required mass range for the $h$--boson, prospects for searching for such objects in the $h\rightarrow b\bar{b}$ and $h\rightarrow \tau\bar{\tau}$ CEP channels would represent an additional advantage of the forward proton approach\footnote{Note however that the situation with the irreducible QCD $b\bar{b}$ CEP background may worsen in the low end of the expected range of the $h$--boson masses. This is due to the fact that in the $b\bar{b}$ CEP channel the signal--to--background ratio in general scales like $\sim{M_{b\
bar{b}}^5}$,\cite{Khoze:2004rc}. Proton tagging with FP420 would also provide a means to search, via the  $h_2\to b\overline{b}$ channel, for the lightest (largely singlet) Higgs boson, $h_2$, of mass $\sim98$ GeV, of the NMSSM scenario~\cite{Belanger:2012tt} where the LHC Higgs--like signal is associated with the heavier Higgs boson, $h_{1}$.}.

Finally we note that the pure pseudoscalar nature of the Higgs boson candidate is already disfavoured by the current data~\cite{DeRoeck,:2012br} and it is not unlikely that by the spring of 2013 the LHC will resolve the fingerprinting issue of the spin and $CP$ parity of the new object, assuming that the latter is conserved in the Higgs sector. However, it will take much more effort to determine whether the observed Higgs--like state has a definite $CP$--parity, and to probe the strength of any possible $CP$--violation, see~\cite{Accomando:2006ga} for a review, in particular Section 3, where the MSSM scenarios with $CP$--violation are discussed. As shown in~\cite{Khoze:2004rc}, the CEP process could become a very promising (and in a sense unique) tool to probe the $CP$--parity of the Higgs--boson candidate. A specific CEP prediction, in the case of a $CP$--violating Higgs boson, is the asymmetry in the azimuthal $\varphi$ distribution of the outgoing protons, caused by the interference of the $CP$--odd and 
$CP$--even vertices in the $gg\to H$ matrix element. The polarisations of the incoming active gluons (see Fig.~\ref{fig:pCp}) are aligned along their respective transverse momenta, and  the contribution caused by the $CP$--odd term in the  $gg\to H$ vertex is proportional to the triple--product correlation
\begin{equation}\nonumber
 \vec{n}_0 \cdot (\vec{p}_{1\perp} \times \vec{p}_{2\perp}) \sim \sin\varphi \;,
\end{equation}
where $\vec{n}_0$ is a unit vector in the beam direction and $\vec{p}_{1,2}$ are the momenta of outgoing protons. In the scenarios discussed in~\cite{Khoze:2004rc,kmrcp} the expected integrated counting asymmetry
\begin{equation}\label{eq:AA}
A=\frac{\sigma(\varphi<\pi)-\sigma(\varphi>\pi)}{\sigma(\varphi<\pi)+\sigma(\varphi>\pi)}\;, 
\end{equation}
could be quite sizeable. Other ideas to search for a direct observation of a $CP$--violating signal in the Higgs sector in CEP are discussed, for instance, in~\cite{JE}.

\section{Conclusion}

To conclude, CEP processes offer a rich phenomenology at high--energy colliders.
 Future CEP data from the LHC and RHIC (as well as 
new analyses from the Tevatron) will undoubtedly
 shed further light on the theory of CEP and exclusive processes.
The installation of forward proton tagging detectors
in the $\sim$200m (AFP and HPS, Stage 1 projects) and 420m regions around ATLAS and/or CMS as well
as the already ongoing promising program
of combined CMS--Totem data taking would certainly add
rich unique capabilities to the existing LHC
experimental studies. We are looking forward to new exciting adventures in Exclusiveland.

\section*{Acknowledgements}
We thank Mike Albrow, Sven Heinemeyer, Risto Orava, Andy Pilkington, Christophe Royon,
Antoni Szczurek, Marek Tasevsky, Misha Voloshin and Georg Weiglein
for  useful discussions. This work was supported by the grant RFBR 11-02-00120-a
and by the Federal Program of the Russian State RSGSS-65751.2010.2. WJS is grateful to the IPPP for an Associateship.\\
VAK thanks the organisers
of the `Diffraction 2012' Workshop
for providing an excellent scientific  environment.

\bibliography{biblio}{}
\bibliographystyle{h-physrev}


\end{document}